\documentclass[final,3p,times,twocolumn]{elsarticle}
\usepackage{graphicx,color}
\usepackage{epsfig}
\usepackage{latexsym,amssymb}

\def\be{\begin{equation}}
\def\ee{\end{equation}}

\def\bc{\begin{center}}
\def\ec{\end{center}}
\def\ds{\displaystyle}
\def\vr{\vec{r}}
\def\vrp{\vec{r'}}

\def\eps{\varepsilon}

\journal{Physica B}

\begin{document}
\begin{frontmatter}

\title{Electron transport in strongly disordered structures.}

\author[rvt]{P. Marko\v{s}}
\address{%
Department of Physics, FEI STU, Ilkovi\v{c}ova 3, 812\,19 Bratislava, Slovakia}
\ead{peter.markos@stuba.sk}

\begin{abstract}
Using the transfer matrix technique, we investigate the propagation of electron through 
a two dimensional disordered sample. We find that
the spatial distribution of electrons is homogeneous only in the limit of weak disorder
(diffusive transport regime).  
In the limit of very strong disorder,  we identify 
a narrow channel through which the   electron propagates from one side of the sample to the
opposite side. Even in this limit, we prove the wave character of the electron propagation.
\end{abstract}


\end{frontmatter}

\section{Introduction}

While the propagation of electrons through weakly disordered samples is 
completely understood \cite{MacKK,Mirlin,DMPK}, the description of electronic transport
in the localized regime still opens a new questions.
Numerically, it was shown \cite{PM,2} that, contrary to the well-established paradigm,
the probability distribution of the logarithm of the conductance is not Gaussian. This  was confirmed by
recent numerical and analytical analysis \cite{prior1,prior3}
and by  analytical formulation of the transport in strongly disordered systems 
\cite{muttalib}.

In Ref. \cite{prior1}, the validity of the single parameter scaling  was confirmed 
numerically in the limit
of strong disorder. Using  the analogy with  statistical polymer models,
the analytical form for the conductance distribution was derived \cite{prior3}.

Muttalib \cite{muttalib} proposed a generalization of the 
Dorokhov Mello Pereira Kumar (DMPK) 
equation \cite{DMPK} to the description of the electron  transport in strongly localized systems. 
Generalized DMPK equation (GDMPK)
contains new  parameters $K_{ab}$, which measure the spatial non-homogeneity of electron  distribution
\cite{MMW}.  Both approximate \cite{MMW} and numerical \cite{BDM}  solutions of GDMPK equation 
agree very well with results of numerical simulations \cite{2}.

Following the the main  idea of GDMPK equation we expect that  due to the strong disorder, 
the spatial  distribution of the electron on the opposite side of the sample is not homogeneous. 
In this paper, we present the  new numerical evidence for  this conjecture.
With the use of the  transfer matrix  numerical analysis,
we study  the spatial distribution of an electron inside the two dimensional 
disordered sample and  show that the  electron distribution is homogeneous   only
in the limit of weak disorder. Stronger disorder causes  the formation of 
continuous cluster of occupied sites inside the sample. This cluster can be 
interpreted as a trajectory
along which electron propagates through the sample. This result agrees with observation of Ref. 
\cite{prior3}. We show that 
the form of this  trajectory is very sensitive to the details of random potential.
and argue that this sensitivity  reflects wave character of the electron propagation
\cite{WaveP}.

\section{Model and  method}\label{model}

The two-dimensional Anderson model \cite{1}  is defined by the Schr\"odinger equation
\be\label{ham}
E\Psi(\vr) 
=  W \epsilon(\vr) \Psi(\vr) +
V\sum_{\vrp}  \Psi(\vrp).
\ee
Electron propagates via hopping from 
the site $\vr$ into the nearest neighbor site $\vrp$, where 
$|\vr-\vrp| = a$  and $a$ is 
the lattice spacing. The size of the system is $L=Na$.
The energies  $\varepsilon(\vr)$ 
are randomly distributed  with the Box 
probability distribution, $P(\epsilon) = 1$ if 
$|\epsilon|<1/2$, and $P(\epsilon)=0$ otherwise.
Also, random energies on different sites are  statistically independent.
The ratio $W/V$ measures  the strength of the disorder. 

\begin{figure}
\bc
\includegraphics[width=6.0cm,clip]{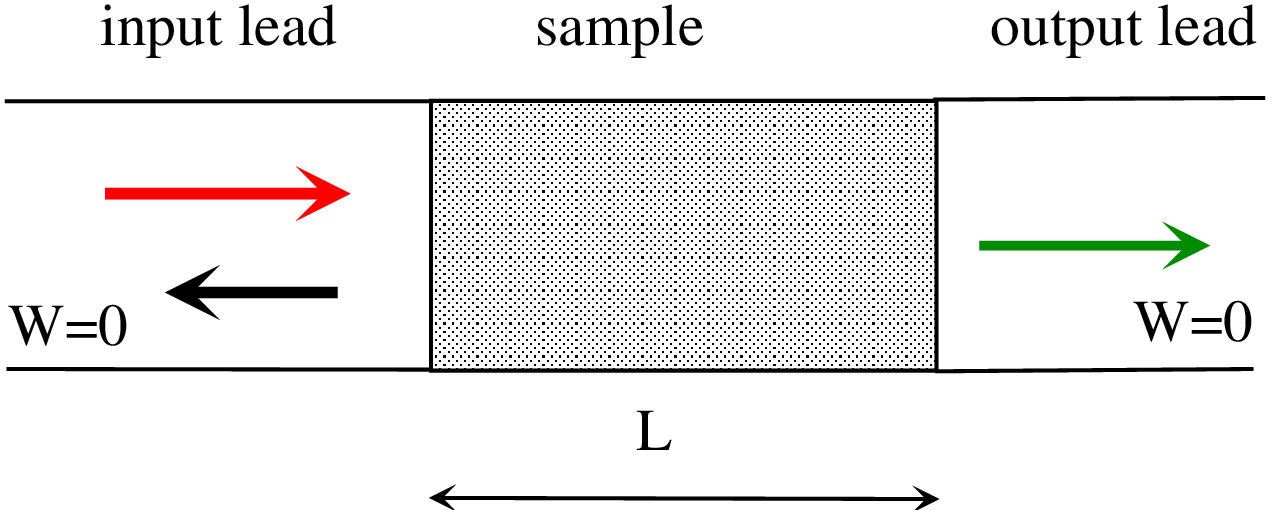}
\ec
\caption{Schematic description of the scattering experiment for the estimation
of the transmission. The sample is connected to two semi-infinite leads, 
represented by tight binding Hamiltonian (\ref{ham}) with zero disorder. 
Electron is coming from the left. It either propagates through the sample
and contributes to the transmission, or  is reflected to the left lead. 
}
\label{obr}
\end{figure}

The disordered sample is connected to 
two semi infinite, disorder free leads which guide 
the electron propagation toward and outward the sample
(Fig. \ref{obr}). The incoming electron either propagates through
the sample, or is reflected back. 
The transmission through the sample is determined by the transfer matrix \cite{PNato}
\be
\textbf{M}=
\left(\matrix{ u & 0  \cr 0 & {u}^* \cr }\right) \left(\matrix{
\sqrt{1+\lambda} & \sqrt{\lambda}   \cr \sqrt{\lambda}   &
\sqrt{1+\lambda} \cr }\right)\left(\matrix{ v & 0  \cr 0 & {v}^*
\cr }\right),
\end{equation}
where $u,v$ are $N \times N$ unitary matrices, and $\lambda$ is a
diagonal matrix with positive elements $\lambda_a, a=1,2, ...N$.

The conductance $g$ is proportional to the transmission $T$   \cite{SE},
\be\label{se}
g = \displaystyle{\frac{e^2}{h}}~T,
~~~~\textrm{and}~~~~T=\sum_a\ds{\frac{1}{1+\lambda_a}}.
\ee
Following  GDMPK,  we expect that  
the probability distribution $P(\{\lambda_a\})$
in the insulating regime 
is influenced  by the distribution of an electron on the opposite side of the sample.
The last is given by parameters $K_{ab}$, defined as
\be
K_{ab}=\sum_{\alpha} |u_{a\alpha}|^2|u_{b\alpha}|^2.
\ee
$K_{ab}=(1+\delta_{ab})/(N+1)$ in the diffusive regime \cite{DMPK}.
However, if the electron distribution is not homogeneous, 
then the matrix elements 
$u_{a\alpha}$ are non-zero only for a small number $n$ of sites ($n\ll N$) and $K_{aa}\sim 1/n\sim 1$. 
\begin{figure}
\bc
\includegraphics[width=5.9cm,clip,angle=-90]{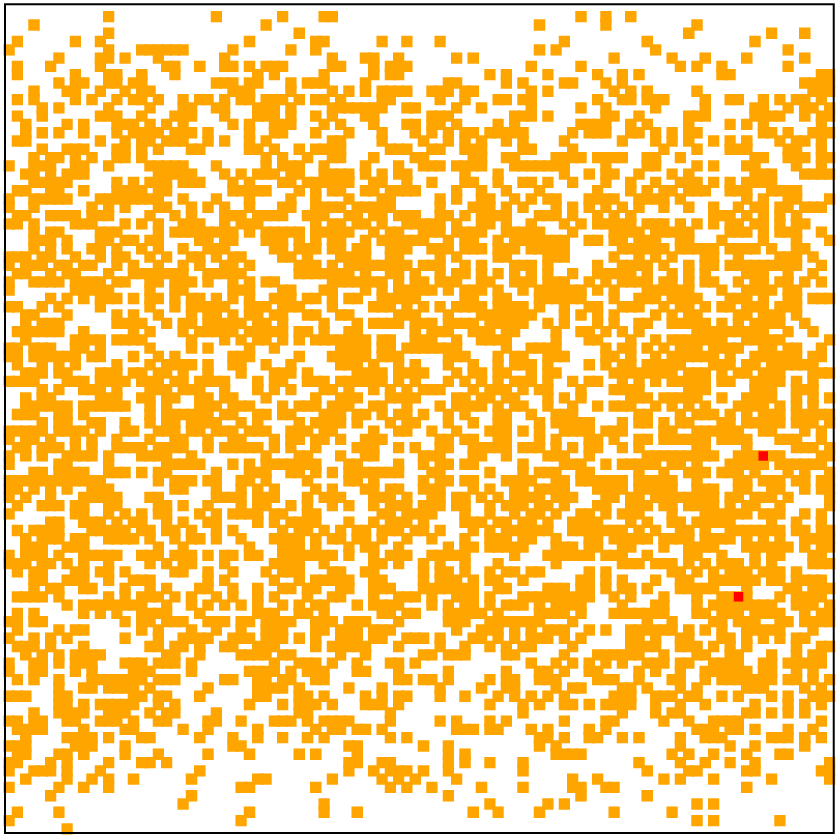}

\vspace*{3mm}
\includegraphics[width=5.9cm,clip,angle=-90]{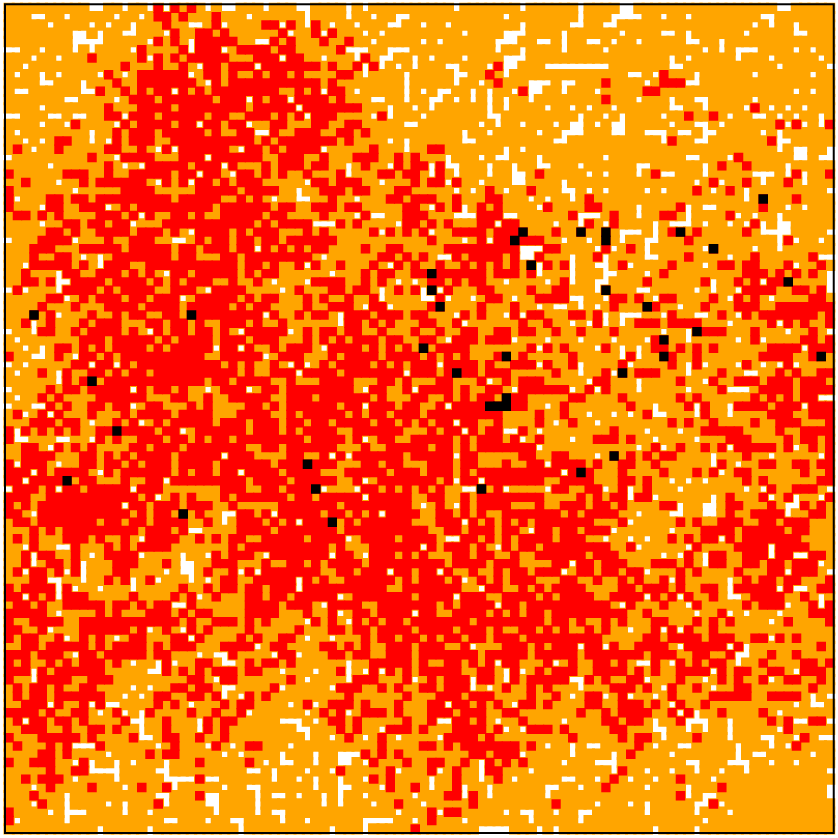}

\vspace*{3mm}
\includegraphics[width=5.9cm,clip,angle=-90]{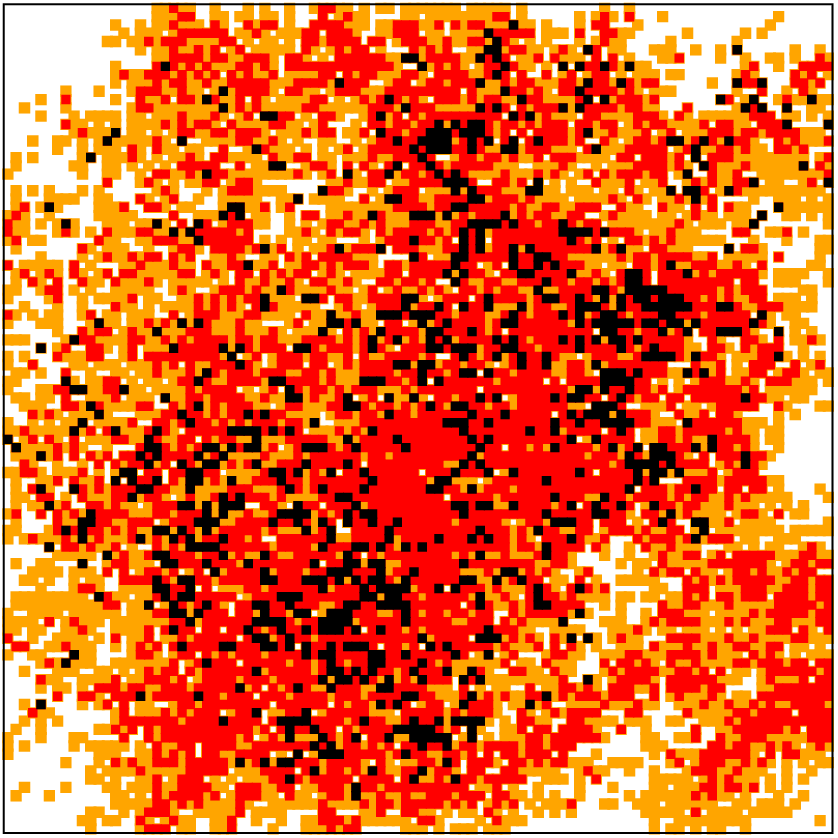}
\ec
\caption{(Color online)
Sensitivity of the transmission trough the disordered system to the change of
the sign of a single random energy $\vr_0$.
Change of the sign of the random energy on orange, red and black sites 
causes the change of the conductance in more than  1\%, 10\% and 100\%, respectively.
The transmission $T_0$  is  $4.998$, $0.52$  and 0.00084 for
the disorder $W/V=2$, 4 and  6 (from top to bottom).
The size of the system is $100a \times 100a$,
and the electron propagates from the left side of the sample to the  right side.
}
\label{w-L100}
\end{figure}
\begin{figure}
\bc
\includegraphics[width=6.4cm,clip,angle=-90]{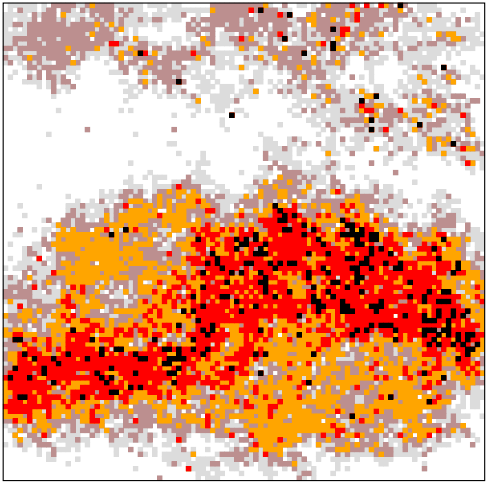}
\ec
\caption{(Color online)
The same as in Fig. \ref{w-L100}  but with disorder $W/V=10$.
The transmission $T_0  = 9\times 10^{-15}$. 
$\langle \ln T \rangle = -33.5$.
Shown are also sites where the change of the sigh  of random energy causes 
the change of the transmission in 0.01\% (gray) and 0.1\% (brown). 
}
\label{w10-L100}
\end{figure}
This conjecture was confirmed by numerical analysis of parameters $K_{ab}$ \cite{MMW} for the three dimensional Anderson model. 
Here, we use the transfer matrix technique 
\cite{Ando-91,PMcKR}  and the  idea of Pichard \cite{PNato},
to visualize  the electron distribution
inside the disordered sample.  
For a given sample, we calculate the transmission $T$, given by Eq. (\ref{se}). 
Then, we create an ensemble of $N^2$ samples, each of them differs from the original one only in the
sign of a single random energy $\varepsilon(\vr)$, and calculate the transmission $T_{\vr}$ for
each sample. 
The relative difference,
\be
\eta(\vr) = \ds{\frac{|T_{\vr}-T|}{T}}
\ee
measures the occupancy of the site $\vr$ \cite{PNato}.
Indeed, $\eta(\vr)$ 
is large only if electron resists at the site $\vr$.
If the wave function $|\Psi(\vr)|$ at site $\vr$ is small,
 then the change of the  random energy $\varepsilon(\vr)$ cannot affect
the transmission $T$  so that $\eta(\vr)$ is small.
The plot of $\eta(\vr)$ enables us to identify the highly occupied sites of a given 
disordered sample.

\begin{figure}
\bc
\includegraphics[width=6.4cm,clip]{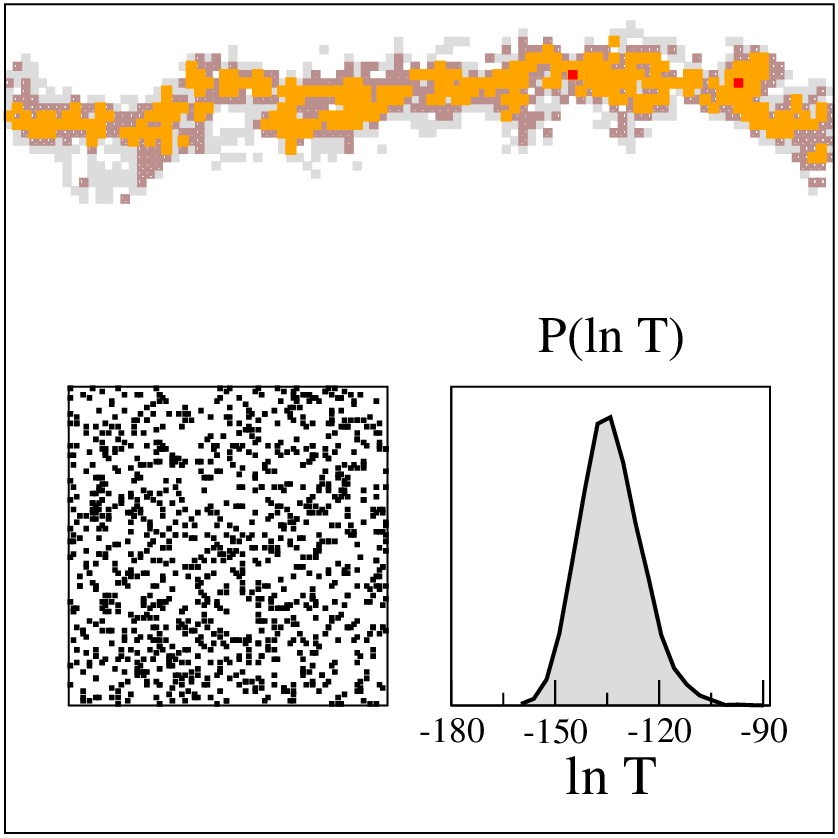}
\ec
\caption{(Color online) The same as in Fig. \ref{w-L100} but for the disorder
$W/V = 20$. 
The transmission of this sample is  $\ln T_0 = -96$.  
Change of the sign of the random energy on 
gray, brown, orange and  red  sites 
causes the change of the logarithm of the 
transmission in more than  0.01\%, 0.1\%, 1\% and  10\%  respectively.
Left inset shows sites where $|\eps(\vr)|<1$. Right inset shows probability 
distribution $P(\ln T)$ with the mean value
$\langle \ln T \rangle = -133$.
}
\label{w20-L100}
\end{figure}

\section{Transmission through disordered sample}\label{path}

\subsection{Weak disorder}
For weak disorder, $W/V=2$ (the localization length $\xi\gg L$),  
the change of only one random energy only  negligibly influences the transmission
Top panel of Fig. \ref{w-L100} shows that the transmission
$T$ changes only in 1\% or even less when the sign of single  random energy $\varepsilon(\vr)$ changes.
Also, it shows that the occupancy of 
all sample sites is  more or  less the same. 
The electron distribution inside the sample is homogeneous,in agreement with the
 DMPK theory  
\cite{DMPK}, and the random matrix theory
of diffusive transport \cite{PNato}.
The lower panels of Fig. \ref{w-L100} demonstrate that the homogeneity of the electron distribution is 
sensitive to the strength of the disorder.

\subsection{Strongly localized limit}

When the disorder increases,
the localization length decreases and becomes smaller than the sample size:
$\xi=5.7a$ ($1.5a$) for disorder $W/V=10$  ($W/V=20$, respectively). 
Although the  typical 
transmission through the strongly disordered sample is  small, we can find,
thanks  to large conductance fluctuations \cite{2}, 
the  sample with relatively large transmission.  

Figures  \ref{w10-L100} and \ref{w20-L100} show that
 the spatial electron distribution is not homogeneous 
inside the  strongly disordered systems.
Some regions of the sample seem not to be occupied. 
With increasing disorder, highly occupied sites  create a continuous  cluster (Fig. \ref{w20-L100}),
which reminds the electron 
trajectory across  the sample \cite{prior3}.  
However,  even in the case of strong disorder
we cannot identify  this cluster  with the  trajectory  known from the classical mechanics.
Indeed, there  are other   sites, randomly distributed in other parts of the sample,
 often located far from the cluster,
which influence the transmission as strongly as the sites on the main cluster
 (Fig. \ref{w10-L100}).
This indicates  that the electron  propagation is 
highly sensitive to any change of the realization of the random potential so that 
the electron wave function is still  distributed throughout the  entire
sample.

The obtained cluster of highly occupied sites  cannot be identified with any 
potential valley or equipotential line  in the 
random potential landscape. To demonstrate this, we show in inset to Fig. \ref{w20-L100} 
that the spatial profile of the random potential does not indicate any  potential valley
in the cluster region. Contrary, as expected for the uncorrelated disorder, 
the spatial  distribution of sites with  random energy $|\varepsilon|<1$ is homogeneous.
We conclude that  the
electron trajectory is the effect   of quantum interference: the electron comes from the left,  
inspect the  sample, and finds the most convenient spatial 
channel to propagate.

\begin{figure}[t!]
\bc
\includegraphics[width=6.4cm,clip,angle=-90]{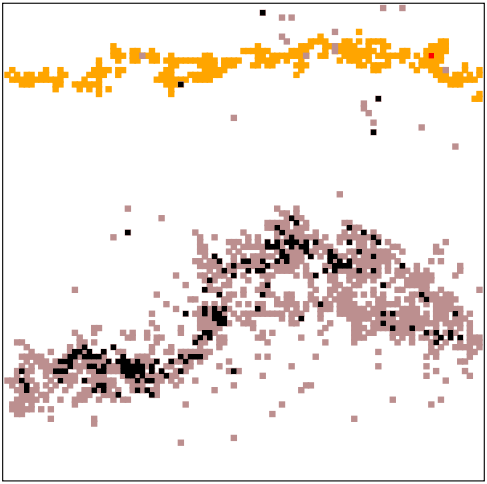}
\ec
\caption{(Color online)
Electron ``trajectory'' through two strongly disordered samples: both samples have the same
realization of random energies. They differ only in the amplitude of fluctuations of random
potential, defined by Eq. (\ref{dva}).
The first sample has $W/V=10$ and the second has $W/V=20$. 
Shown are the lattice sites where the change of the sign of random energy affects the 
change of the logarithm of the transmission  in 1\% and 10\%.
The electron prefers 
completely different trajectories through these samples. Brown and black sites represent
the path of the electron for $W/V=10$, orange and red sites show the path for $W/V=20$.
}
\label{w10-w20}
\end{figure}

To support this claim,  we consider two disordered samples, which 
have the same potential profile  but 
differ  in amplitude of random fluctuations: $W/V=10$ for the sample I
and $W/V = 20$ for the sample II:
\be\label{dva}
\varepsilon(\vr)^{II} = 2\varepsilon(\vr)^{I}
\ee
for all lattice sites $\vr$. With the use of the above mentioned method, we calculate
the highly occupied sites for both samples.
There is no reason to expect that the position of these sites  changes in  the case 
of classical particle.
However, as shown in Fig. \ref{w10-w20},
the electron chooses  completely different trajectories 
through the two samples.
The increase  of fluctuations of the random potential
causes that electron prefers to transmit through completely different sites than
it was in the sample I.

\section{Conclusion}

We described the propagation of quantum particle through
a disordered sample and show how this propagation depends on the strength  of the
disorder. 
Our data confirm that the distribution of the electron inside the sample is  homogeneous
only when the disorder is small.
In the limit of strong localization, we  find a continuous cluster of  preferably visited sites
which can be interpreted as a 
electron trajectory through the sample.
This result is consistent with the recent
model for the transport through the insulators \cite{prior3}. 
We also proved  that  the obtained  trajectory 
does not contradict the  quantum character of electron
propagation, and cannot be identify with the
trajectory of the classical particle propagating through the sample. 

\bigskip

This work was supported by project APVV n. 51-003505  and project VEGA 0633/09.

\end{document}